\documentclass[aps,pre,twocolumn]{revtex4-1}
\usepackage{hyperref}
\usepackage{color}
\usepackage{graphics,graphicx,epsfig,wrapfig}
\usepackage{epsf,epstopdf}
\usepackage{amssymb,amsfonts,amsmath}
\usepackage{enumitem}

\usepackage{ifthen}

\newcommand\mydots{\hbox to 0.8em{.\hss.\hss.}}
\newcommand{\beqn}{\begin{eqnarray}}
\newcommand{\eeqn}{\end{eqnarray}}
\newcommand{\beq}{\begin{equation}}
\newcommand{\eeq}{\end{equation}}

\DeclareMathOperator*{\argmax}{arg\,max}

\begin{document}

\title{Method for identification of condition-associated public antigen receptor sequences}
\author{M.V. Pogorelyy$^1$, A.A. Minervina$^1$, D.M. Chudakov$^{1-5}$, I.Z. Mamedov$^{1-3}$, Y.B. Lebedev$^{1,6*}$,\\ T. Mora$^{7*}$, A.M. Walczak$^{8*}$}
\affiliation{~\\
$^1$ Department of genomics of adaptive immunity, IBCH RAS, Russia\\
$^2$ Center for Data-Intensive Biomedicine and Biotechnology, Skoltech, Russia\\
$^3$ Genomics of antitumor adaptive immunity laboratory, Nizhny Novgorod State Medical Academy, Russia\\
$^4$ Department of molecular technologies, Pirogov Russian National Research Medical University, Russia\\
$^5$ Central European Institute of Technology, CEITEC, Czech republic\\
$^6$ Biological faculty, Moscow State University, Russia\\
$^7$ Laboratoire de physique statistique, CNRS, UPMC and \'Ecole normale sup\'erieure, Paris, France\\
$^8$ Laboratoire de physique th\'eorique, CNRS, UPMC and \'Ecole normale sup\'erieure, Paris, France\\
* \textrm{\texttt{lebedev\_yb@mx.ibch.ru, tmora@lps.ens.fr, awalczak@lpt.ens.fr}}
}

\begin{abstract}
Diverse repertoires of hypervariable immunoglobulin receptors (TCR and BCR) recognize antigens in the adaptive immune system. The development of immunoglobulin receptor repertoire sequencing methods makes it possible to perform repertoire-wide disease association studies of antigen receptor sequences. We developed a statistical framework for associating receptors to disease from only a small cohort of patients, with no need for a control cohort. Our method successfully identifies previously validated Cytomegalovirus and type 1 diabetes responsive receptors.  
\end{abstract}

\maketitle

T-cell receptors (TCR) and B-cell receptors (BCR) are hypervariable immunoglobulins that play a key role in recognizing antigens in the vertebrate immune system. TCR and BCR are formed in the stochastic process of V(D)J recombination, creating a diverse sequence repertoire. Progress in high throughput sequencing now allows for deep profiling of T-cell repertoires, by establishing a near-complete list of unique receptor sequences, or ``clonotypes,'' present in a sample.

Comparison of sequenced repertoires has revealed that in any pair of individuals, large numbers of TCR sequences have the same amino acid sequence \cite{Venturi2011}. Several mechanisms leading to the repertoire overlap have been identified so far. The first mechanism is {\em convergent recombination}. Due to biases in V(D)J recombination process, the probability of generation of some receptors is very high, making them appear in almost every individual multiple times and  repeatedly sampled in repertoire profiling experiments \cite{Britanova2014}. This sharing does not result from a common specificity or function of the shared clonotypes, and may in fact correspond to cells from the naive compartment in both donors\cite{Quigley2010}, or from functionally distinct subsets such as CD4 and CD8 T-cells. The second possible reason for TCR sequence sharing is specific to identical twins, who may share T cell clones as a consequence of cord blood exchange {\em in utero} via a shared placenta \cite{Pogorelyy2017}. The third and most interesting mechanism for sharing receptor sequences is {\em convergent selection}, in response to a common antigen. From functional studies, such as sequencing of MHC-multimer specific T-cells, it is known that the antigen-specific repertoire is often biased, and the same antigen-specific TCR sequences can be found in different individuals \cite{Miles2011,Dash2017,Glanville2017}. 

\begin{figure}
\includegraphics[width=\linewidth]{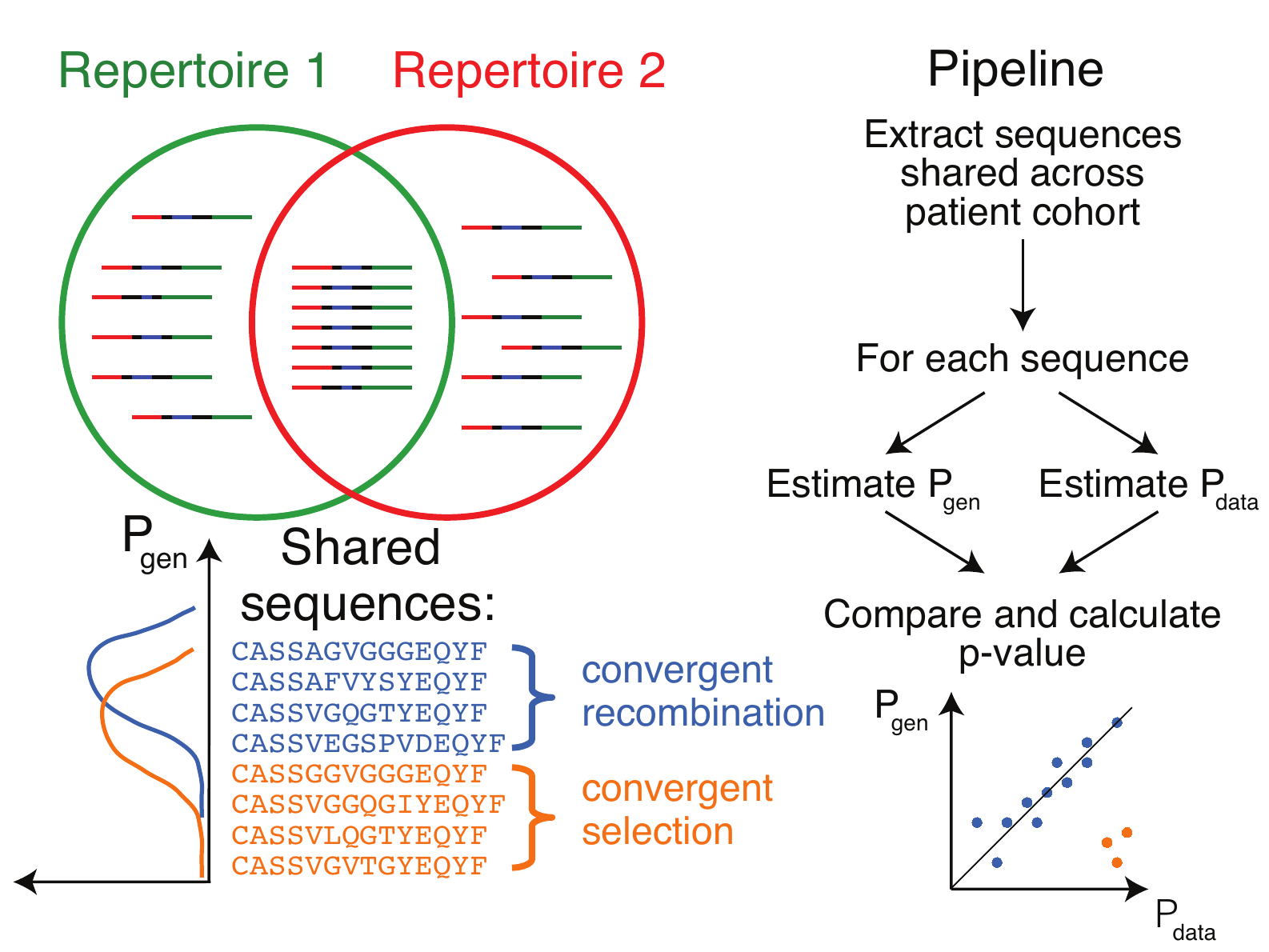}
\centering
\caption{\textbf{Method principle and pipeline}. (Top left) Sequence overlap between two TCR or BCR repertoires. (Bottom left) There are two major mechanisms for sequence sharing between two repertoire: convergent recombination and convergent selection. Because convergent recombination favors sequences with high generation probability, these two classes of sequences have different distribution of generative probability, $P_{\rm gen}(\sigma)$. (Right) We estimate the theoretical $P_{\rm gen}(\sigma)$ for each sequence $\sigma$ and compare it to $P_{\rm data}(\sigma)$, which is empirically derived from sharing pattern of that sequence in the cohort. Comparison of these two values allows us to calculate the analog of a p-value, namely the posterior probability that the sharing pattern is explained by convergent recombination alone, with no selection for a common antigen.
}\label{fig:cartoon}
\end{figure}

Reproducibility of a portion of the antigen-specific T-cell repertoire in different patients creates an opportunity for disease association studies using T-cell repertoire datasets \cite{Faham2017,Emerson2017}. These studies analyse the TCR sequence overlap in large cohorts of healthy controls and patients to identify shared sequences overrepresented in the patient cohort. Here we propose a novel computational method to identify clonotypes which are likely to be shared because of selection for their response to a common antigen, instead of convergent recombination. Our approach is based on a mechanistic model of TCR recombination and is applicable to small cohorts of patients, without the need for a healthy control cohort.  

As a proof of concept, we applied our method to two large publicly available TCR beta datasets from Cytomegalovirus (CMV)-positive \cite{Emerson2017} and type 1 diabetes (T1D) \cite{Seay2016} patients. In both studies the authors found shared public TCR clonotypes that are specific to CMV-peptides or self-peptides, respectively. Specificity of these clonotypes was defined using MHC-multimers. We show that clonotypes functionally associated with CMV and T1D in these studies are identified as outliers by our method. 

\begin{figure*}
\includegraphics[width=\linewidth]{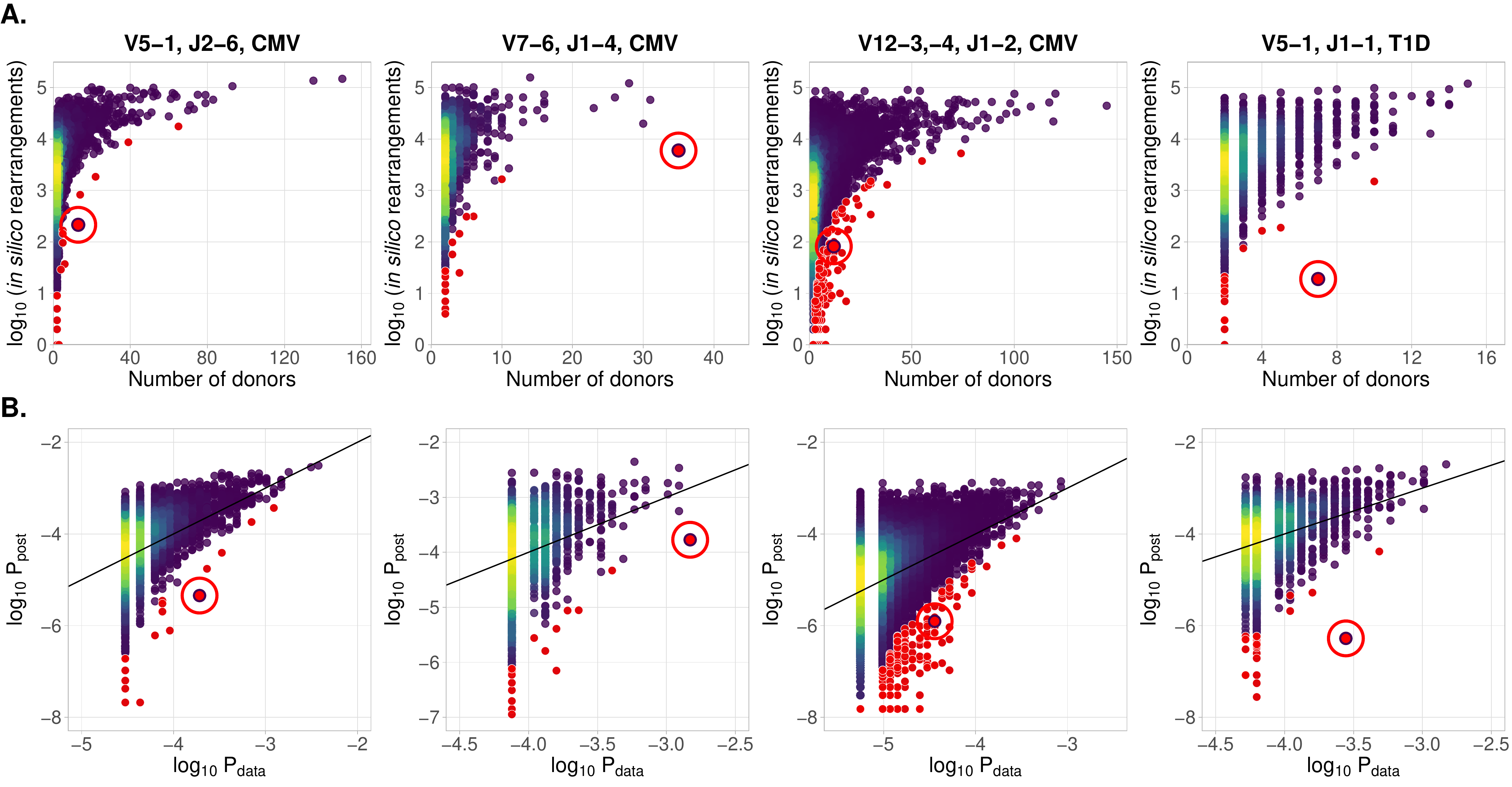}
\caption{\textbf{A. CDR3aa of antigen specific clonotypes (red circles) have less generative probability than other clonotypes shared among the same number of donors.} Generation probability for each clonotype in a given VJ combination, $P_{\rm post}(\sigma)$, plotted against the number of patients with that clonotype. 
\textbf{B. Model prediction of generative probability agrees well with data.} To directly compare $P_{\rm post}(\sigma)$ to data, we estimate the empirical probability of occurrence of sequences, $P_{\rm data}(\sigma)$, from its sharing pattern across donors (see Methods). In A. and B. red dots indicate significant results (adjusted $P<0.01$, Holm's multiple testing correction), while red circles point to the responsive clonotypes identified in the source studies. 
}\label{fig:plots}
\end{figure*}

The main ingredient of our approach is to estimate the probability of generation of shared clonotypes, and to use this probability to determine the source of sharing (see Fig. \ref{fig:cartoon}). Due to the limited sampling depth of any TCR sequencing experiment, chances to sample the same clonotype twice are low, unless this clonotype is easy to generate convergently, with many independent generation events with the same TCR amino acid sequence in each individual (convergent recombination), or if this clonotype underwent clonal expansion, making its concentration in blood high (convergent selection).
Thus, we reasoned that convergently selected clonotypes should have a {\em lower} generative probability than typical convergently recombined clonotypes. To test this, we estimated the generative probability of the TCR's beta chain Complementarity Determining Region 3 (CDR3) amino-acid sequences that were shared between several patients. Since no algorithm exists that can compute this generative probability directly, our method relies on the random generation and translation of massive numbers of TCR nucleotide sequences using a mechanistic statistical model of V(D)J recombination \cite{Murugan2012}, as can be easily performed e.g. using the IGoR software \cite{Marcou2017}.

In Fig.\ref{fig:plots}A we plot for each clonotype the number of donors sharing that clonotype against its generation probability. Disease-specific TCR variants validated by functional tests in source studies are circled in red.  Note that validated disease-specific TCRs have a much lower generation probability than the typical sequences shared by the same number of donors. We developed a method of axis transformation (see SI) to compare the model prediction with data values on the same scale (Fig.\ref{fig:plots}B), so that outliers can be easily identified by their distance to identity line. 
Our method can be used to narrow down the potential candidates for further experimental validation of responsive receptors.

Our method also identifies other significant outliers than reported in the source studies (shown in red, and obtained after multiple-test correction -- see Methods), which may have three possible origins. First, they may be associated with the condition, but were missed by the source studies.
Second, they may be due to other factors shared by the patients, such as features involved in thymic or peripheral selection, or reactivity to other common conditions than CMV (e.g. influenza infection). Third, they can be the result of intersample contamination. Our approach is able to diagnose the last explanation by estimating the likelihood of sharing at the level of nucleotide sequences (i.e. synonymously), as detailed in the Methods section.

Our approach can be used on other hypervariable receptor chains (TCR alpha, BCR heavy and light chains), as well as other species (mice, fish, etc.). Recent advances in computational methods allow us to extract TCR repertoires from existing RNA-seq data \cite{Bolotin2017,Brown2015}.
Huge numbers of available RNAseq datasets from patients with various conditions can be used for analysis and identification of novel virus, cancer, and self reactive TCR variants using our method. The more immunoglobulin receptors with known specificity are found using this type of association mapping, the more clinically relevant information can be extracted from immunoglobulin repertoire data.

\section*{Acknowledgments}
This work was supported by Russian Science Foundation grant 15-15-00178, and  and partially supported by European Research Council Starting Grant 306312. 

\section*{Methods}
\setcounter{section}{1}
\subsection{Statistical analysis}
\subsubsection{Problem formulation}
Our framework is applicable to analyze the outcome of a next generation sequencing experiment probing the immune receptor repertoires of $n$ individuals with a given condition, e.g. CMV or Type 1 diabetes. We denote by $M_i$ the number of unique amino acid TCR sequences in patient $i$, $i=1,\ldots,N$. For a given TCR amino acid sequence $\sigma$, we set $x_i=1$ to indicate that $\sigma$ is present in patient $i$'s repertoire, and $x_i=0$ otherwise.
For a given shared sequence $\sigma$, we want to know how likely its sharing pattern is under the null hypothesis of convergent recombination, correcting for the donors' different sampling depths. In other words, is $\sigma$ overrepresented in the population of interest? If $\sigma$ is significantly overrepresented, we also want to quantify the size of this effect. 

\subsubsection{Overview}

Under the null hypothesis, the presence of $\sigma$ in a certain number of donors is explained by independent convergent V(D)J recombination events in each donor.
Given the total number of recombination events that led to the sequenced sample of donor $i$, $N_i$, the presence of given amino acid sequence $\sigma$ in donor is Bernoulli distributed with probability 
\begin{align}
p_i=\langle x_i\rangle &=\left(1-P_{\rm post}(\sigma)\right)^{N_i},\\ 
P_{\rm post}(\sigma)&=P_{\rm gen}(\sigma) \times Q,
\end{align}
where $P_{\rm post}(\sigma)$ is the model probability that a recombined product found in a blood sample has sequence $\sigma$ under the null hypothesis. It is formed by the product of
$P_{\rm gen}(\sigma)$, the probability to generate the sequence $\sigma$, estimated using a V(D)J recombination model (see \ref{sec:pgen}), and $Q$, a constant correction factor accounting for thymic selection (see \ref{sec:qfac}). The number of independent recombination events $N_i$ leading to the observed unique sequences in a sample $i$ is unknown, because of convergent recombination events within the sample, but it can be estimated from the number of unique sequences $M_i$, using the model distribution $P_{\rm post}$ (see \ref{sec:ni}).

We also calculate the posterior distribution of $P_{\rm data}(\sigma)$, corresponding to 
the empirical counterpart of $P_{\rm post}(\sigma)$ in the cohort, inferred from the sharing pattern of $\sigma$ across donors. We use information about the presence of $\sigma$ in our donors, $x_1,\ldots,x_n$ and the sequencing depth for each donor, $N_1,\ldots,N_n$  (see \ref{sec:pdata}), yielding the posterior density: $\rho(P_{\rm data}|x_1,\ldots,x_N)$.

Finally, we estimate the probability, given the observations, that the true value of $P_{\rm data}$ is smaller than the theoretical value $P_{\rm post}$ predicted using V(D)J recombination model, analogous to a p-value and used to identify significant effects:
\begin{equation}\label{eq:pvalue}
\mathbb{P}(P_{\rm post}>P_{\rm data})=\int_{0}^{P_{\rm post}}\rho(P_{\rm data}|x_1,\ldots,x_n) dP_{\rm data}.
\end{equation}

To estimate the effect size $q(\sigma)$ we compare $P_{\rm data}$ to $P_{\rm post}$,
\begin{equation}\label{eq:effectsize}
q(\sigma)=\frac{P_{\rm data}(\sigma)}{P_{\rm post}(\sigma)}.
\end{equation}

\subsubsection{Estimation of $P_{\rm gen}$, the probability of generation of a TCR CDR3 amino acid sequence} \label{sec:pgen}
To procedure outlined above requires to calculate $P_{\rm gen}(\sigma)$, the probability to generate a given CDR3 amino acid sequence. Methods exist to calculate the probability of TCR and BCR nucleotide sequences from a given recombination model \cite{Murugan2012,Marcou2017}, but are impractical to calculate the probability of amino acid sequences, because of the large number of codon combinations that can lead to the same amino acid sequence, $\prod_{a=1}^{L}{n_{\rm codons}(\sigma(a))}$, where $L$ is the sequence length, and $n_{\rm codons}(\tau)$ the number of codons coding for amino acid $\tau$. The number is about $1.4\times 10^7$ for a typical CDR3 length of 15 amino acid.

Instead, we estimated $P_{\rm gen}(\sigma)$ using a simple Monte-Carlo approach. We randomly generated a massive number ($N_{\rm sim}=2\times 10^9$) of recombination scenarios according to the validated recombination model \cite{Murugan2012}:
\begin{eqnarray}
P_\text{rearr}^\beta (r) &=& P(V)P(D,J) P(\text{del}V|V) P(\text{ins}VD)   \\
&& \times P(\text{del}Dl,\text{del}Dr|D) P(\text{ins}DJ) P(\text{del}J|J) \nonumber.
\end{eqnarray}
The resulting sequences were translated, truncated to only keep the CDR3, and counted. $P_{\rm gen}(\sigma)$ was approximated by the fraction of events thus generated that led to sequence $\sigma$. This approximation becomes more accurate as $N_{\rm sim}$ increases, with an error on $P_{\rm gen}(\sigma)$ scaling as $(P_{\rm gen}(\sigma)/N_{\rm sim})^{1/2}$.

\subsubsection{Estimation of the correction factor $Q$}\label{sec:qfac}
Not all generated sequences pass selection in the thymus. $P_{\rm gen}$ systematically underestimates the frequency of recombination event that eventually make it into the observed repertoire. To correct for this effect, we estimate a correction factor $Q$, as was suggested in \cite{Elhanati2014}: 
\begin{equation}
P_{\rm post}(\sigma)=P_{\rm gen}(\sigma)\times Q.
\end{equation}
Contrary to \cite{Elhanati2014}, which learned a sequence-specific factor for each individual, here
we assume that all observed sequences passed thymic selection. $Q$ is a normalization factor accounting for the fact that just a fraction $Q^{-1}$ of sequences pass thymic selection.
This factor is determined for each VJ-combination as an offset when plotting $\log{P_{\rm gen}}$ against $\log{P^*_{\rm data}}$ (see \ref{sec:pdata} for definition of $P^*_{\rm data}$),  using least squares fitting.

\subsubsection{Estimation of $P_{\rm data}(\sigma)$, the probability of sequence occurrence in data}\label{sec:pdata}
The variable $x_i$ indicates the presence or absence of a given TCR amino acid sequence $\sigma$ in  the $i$th dataset with $N_i$ recombination events per donor. We want to estimate $P_{\rm data}(\sigma)$, which is a fraction of recombination events leading to $\sigma$ in the population of interest. 
According to Bayes' theorem, for a given $\sigma$, the probability density function of $P_{\rm data}$ reads: 
\begin{widetext}
\begin{equation}
\rho(P_{\rm data} | x_1,\ldots,x_n)=\frac{\mathbb{P}(x_1,\ldots,x_n | P_{\rm data})\rho_{\rm prior}(P_{\rm data})}{\int_{0}^{1}{\mathbb{P}(x_1,\ldots,x_n | P_{\rm data})\rho_{\rm prior}(P_{\rm data})\,dP_{\rm data}}}.
\end{equation}
The likelihood is given by a product of Bernouilli probabilities: 
\begin{equation}
\mathbb{P}(x_1,\ldots,x_n | P_{\rm data})=\prod_{i=1}^N \left[1-(1-P_{\rm data})^{N_i}\right]^{x_i}\left[(1-P_{\rm data})^{N_i}\right]^{1-x_i},
\end{equation}
\end{widetext}
and a flat prior $\rho_{\rm prior}(P_{\rm data})={\rm const}$ is used. 

We estimate $P^*_{\rm data}$ (shown in Fig. 2B) as the  maximum of the posterior distribution:
\begin{equation}
P^*_{\rm data}=\argmax_{P_{\rm data}}\rho(P_{\rm data} | x_1,\ldots,x_n).
\end{equation}

\subsubsection{Estimation of $N_i$, the number of recombination events}\label{sec:ni}
The total number $N_i$ of recombination events in $i$th dataset is unknown, but we can count the number of unique CD3 acid sequences $M_i$ observed in the sequencing experiment. For a typical TRB experiment, convergent recombination is relatively rare and one could use $N_i\approx M_i$ as an approximation. However, for less diverse loci (e.g TRA), or for much higher sequencing depths, one should correct for convergent recombination, as the the observed number of unique aminoacid sequences could be much lower than the actual number of corresponding recombination events.

The average number of unique sequences resulting from $N_i$ recombination events is, in theory:
\begin{equation}\label{eq:calibration}
\langle M_i\rangle = \sum_{\sigma\in T} (1-P_{\rm post}(\sigma))^{N_i}.
\end{equation}
where $T$ is the set of sequences that can pass thymic selection. To estimate that number, we generate a very large number $N_{\rm sim}$ of recombinations, leading to $N_{\rm uni}$ unique CDR3 amino acid sequences for which $P_{\rm gen}$ is estimated as explained above. We take $T$ to be a random subset of unique sequences, $T\subset \{\sigma_1,\ldots,\sigma_{N_{\rm uni}}\}$, of size $|T|=N_{\rm uni}/Q$, and we apply Eq.~\ref{eq:calibration}.

\begin{figure}[h]
\centering
\includegraphics[width=\linewidth]{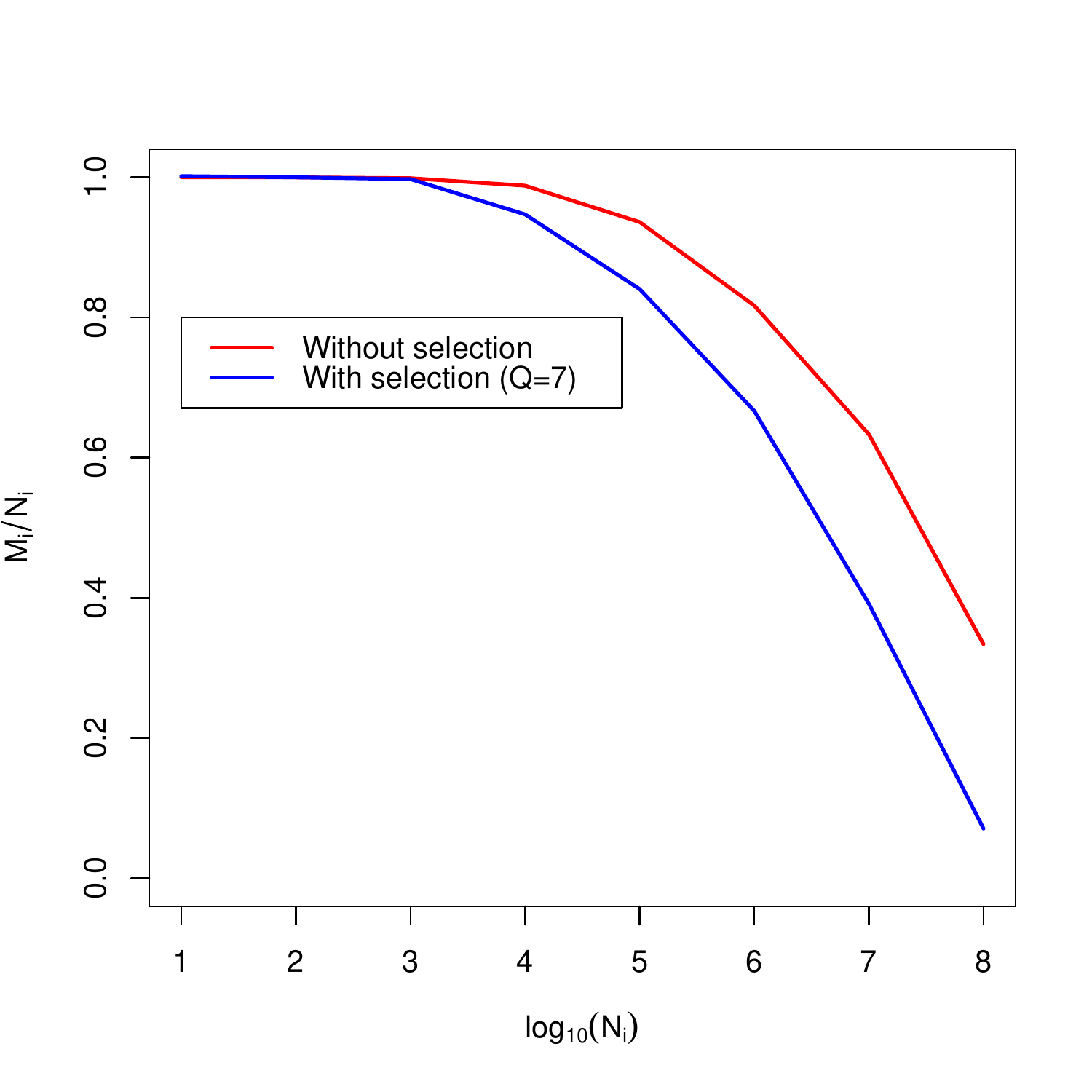}
\caption{\textbf{Calibration curve for TRBV5-1 TRBJ2-6 combination.} Here we plot fraction of unique aminoacid sequences to recombination events against logarithm of recombination events number. Blue line corresponds to theoretical solution with selection, red line corresponds to theoretical solution without selection. 
\label{figcalibration}}
\end{figure}

Using this equation we plot the calibration curve for the TRBV5-1 TRBJ2-6 VJ datasets in Fig.~\ref{figcalibration}. For comparison the case of no thymic selection ($Q=1$) is shown in red. The inversion of this curve yields $N_i$ as a function of $M_i$.

\subsection{Pipeline description}\label{sec:pipeline}
In this section we describe how to apply our algorithm to real data. All the code and data necessary to reproduce our analysis is available online on github (\url{https://github.com/pogorely/vdjRec/}).

We start with annotated TCR datasets (CDR3 amino acid sequence, V-segment, J-segment), one per donor. Such datasets are produced by  MiXCR\cite{Bolotin2015}, immunoseq (\url{http://www.adaptivebiotech.com/immunoseq}) and most other software for NGS repertoire data preprocessing. Data we used was in immunoseq format, publicly available from \url{https://clients.adaptivebiotech.com/immuneaccess} database.

We proceed as follows:
\begin{enumerate}

\item{Split datasets by VJ combinations. The resulting datasets correspond to lists of unique CDR3 amino acid sequences for each donor and VJ combination.
All following steps should be done independently for each VJ combination.}
\item{(Optional). Filter out sequences present in only one donor to speed up the downstream analysis.}
\item{Generate a large amount of simulated nucleotide TCR sequences for a given VJ combination. Extract and translate their CDR3, and count how many times each sequence appears in the simulated set (restricting to sequences actually observed in donors for better efficiency). The resulting number divided by the total number of simulated sequences is an estimate of $P_{\rm gen}$.}
\item{Estimate $P^*_{\rm data}$ for each sequence in the dataset, see \ref{sec:pdata}}.
\item{Using $P^*_{\rm data}$ and $P_{\rm gen}$, estimate for each VJ combination the normalization $Q$ by minimizing $\sum_{j=1}^{n} (\log P^*_{\rm data}(\sigma_j)-\log P_{\rm gen}(\sigma_j)-\log Q)^2$, see \ref{sec:qfac}}, where $\sigma_j$, $j=1,\ldots,n$ are the shared sequences.
\item{Calculate $P_{\rm post}=Q \times P_{\rm gen}$. Calculate the p-value (Eq.~\ref{eq:pvalue}) and effect size (Eq.~\ref{eq:effectsize}).} 
\end{enumerate}

\subsection{Usage example}
\subsubsection{Data sources}
Data from \cite{Emerson2017} and \cite{Seay2016} is publicly available from the immuneaccess database: \url{https://clients.adaptivebiotech.com/immuneaccess}. For our analysis, we only considered VJ combinations for which the authors identified condition-associated clonotypes with MHC-multimer proved specificity. CDR3 aminoacid sequences and V and J segment of these TCR clonotypes are given in Table \ref{tbl:clones}. 

\begin{table*}[!ht]
\begin{center}
\begin{tabular}{ l| l| l| l| l }
\hline
CDR3aa & V-segment & J-segment & Antigen source& Ref.  \\
\hline 
\texttt{CASSLAPGATNEKLFF}&TRBV07-06&TRBJ1-4&CMV&\cite{Emerson2017} \\
\hline 
\texttt{CASSPGQEAGANVLTF}&TRBV05-01&TRBJ2-6&CMV&\cite{Emerson2017} \\
\hline 
\texttt{CASASANYGYTF}&TRBV12-3,-4&TRBJ1-2&CMV&\cite{Emerson2017} \\
\hline 
\texttt{CASSLVGGPSSEAFF}&TRBV05-01&TRBJ1-1&self&\cite{Seay2016,Gebe2009}\\
\end{tabular}
\caption{Published antigen-specific clonotypes used to test the algorithm.\label{tbl:clones}}
\end{center}
\end{table*}
\begin{table*}[!ht]\label{tbl:results}
\begin{center}
\begin{tabular}{ l| l| l| l| l| l |l| l }
\hline
CDR3aa & V & J & Ag.source & Ref.&p-value rank& p-value& Effect size  \\
\hline 
\texttt{CASSLAPGATNEKLFF}&07-06&1-4&CMV&\cite{Emerson2017}&1/1637&$1.2 \times 10^{-17}$&8.8 \\
\hline 
\texttt{CASSPGQEAGANVLTF}&5-01&2-6&CMV&\cite{Emerson2017}&1/5549&$1.8 \times 10^{-17}$&42.3 \\
\hline 
\texttt{CASASANYGYTF}&12-3,-4&1-2&CMV&\cite{Emerson2017}&40/27669&$2.5 \times 10^{-14}$&28.8 \\
\hline 
\texttt{CASSLVGGPSSEAFF}&5-01&1-1&self&\cite{Seay2016,Gebe2009}&1/2646& $9.5\times 10^{-19}$ &524\\
\end{tabular}
\caption{Output of the algorithm for sequences from table \ref{tbl:clones}.}
\end{center}
\end{table*}

\subsubsection{Analysis results}
We applied our pipeline to identify CMV-specific and self-specific TCR sequences listed in Table \ref{tbl:clones}. 
For our analysis we used only case cohorts, without controls.
For each dataset we 
followed our pipeline described in \ref{sec:pipeline}. We found that sequences reported in the source studies as being both significantly enriched in the patient cohort, and antigen-specific according to MHC-multimers, were the most significant in  3 out of 4 datasets. In the remaining TRBV12 dataset, the sequence of interest was the top $40$ most significant out of $27,699$ sequences present in at least two CMV-positive donors.

\subsection{Identifying contaminations}\label{sec:contam}
Intersample contamination may complicate high-throughput sequencing data analysis in many ways. It could occur both during library preparation or the sequencing process itself \cite{Sinha125724}. Contaminations have the same nucleotide and amino acid sequence in all datasets, and so our method identifies them as outliers, because their sharing cannot be explained by a high recombination probability. 

Our method provides a tool to diagnose contamination. Given an amino-acid sequence present in many donors, we measure its theoretical nucleotide diversity using the same simulation approach we used to calculate the generative probability $P_{\rm gen}$ of the amino acid sequence (see \ref{sec:pgen}). 
If the diversity of the simulated nucleotide sequences is much larger than observed in the data, it is a sign of contamination.  

We applied this approach to the CDR3 sequence \texttt{CASSLVGGPSSEAFF} associated to Type 1 diabetes, and found 19 recombination events consistent with that amino acid sequence out of our simulated dataset. We found 18 different nucleotide variants out of the 19 total possible. In contrast, in the data this clononotype had the same nucleotide variant in all of the 8 donors in which it was present. That variant was absent from the simulated set. 
A one-sided Fisher exact test gives a $p<10^{-6}$ probability of this happening by chance, indicating contamination as a likely source of sharing.

\bibliographystyle{pnas}

\end{document}